# Nonlinear All Optical Digital Amplification of the Light Pulse in Weakly Coupled Photonic Crystal Waveguides

Vakhtang Jandieri, *Senior Member, IEEE* and Ramaz Khomeriki

*Abstract* – **All-optical amplification of the light pulse in a weakly coupled two nonlinear photonic crystal waveguides (PCWs) is proposed. We consider pillar-type PCWs, which consist of the periodically distributed circular rods made from a Kerr-type dielectric material. Dispersion diagrams of the symmetric and antisymmetric modes are calculated. The operating frequency is properly chosen to be located at the edge of the dispersion diagram of the modes. In the linear case no propagation modes are excited at this frequency, however, in case of nonlinear medium when the amplitude of the injected signal is above some threshold value, the solitons are formed and they are propagating inside the coupled nonlinear PCWs. Near field distributions of the light pulse propagation inside the coupled nonlinear PCWs and the output powers of the registered signals are studied in a detail. The amplification coefficient is calculated at the various amplitudes of the launched signal. The results vividly demonstrate the effectiveness of the weakly coupled nonlinear PCWs as all-optical digital amplifier.**

## I. INTRODUCTION

PHOTONIC crystals have inspired a lot of interest due to their wide application for controlling of light [1,2].

Photonic crystals are artificial dielectric or metallic structures in which any electromagnetic wave propagation is forbidden within a fairly large frequency range. Photonic crystals have been used to replace the usual dielectric or metallic elements of the devices to improve their electromagnetic properties. Where the optical response remains linear, the photonic crystals have been successfully used to design the ultra-compact and miniaturized photonic devices. However, for the realization of true all-optical signal processing the optical system needs to carry nonlinear properties. In optically nonlinear media the index of refraction is modified by the presence of a light signal and this modification can be exploited to influence another light signal, thereby performing an all-optical signal processing operation.

Photonic crystals are the most promising candidate to enable optical integration. Integration of many functions on the same chip leads to much lower production and operating costs. Recently, a planar silicon PCW has enabled the successful realization of various nonlinear optical devices on a chip. PCW consists of a line defect introduced in a perfect photonic crystals and they are the unique structures due to their ability for strong light confinement [3,4]. Various interesting nonlinear phenomena in silicon PCW, such as slow light enhancement of the optical nonlinear effects [5] and soliton-effect pulse compression of picosecond pulse [6], have been observed. All of these features are technologically very important and could lead as to the deeper understanding of the nonlinear wave propagation in the silicon medium, as well as to the creation of the ultra-fast, all-optical signal processing devices that are miniature and suitable for integration.

In our recent papers we have proposed the realistic model of all-optical amplifier using subwavelength metallic waveguides [7] or the coupled linear PCWs [8]. The effect of all-optical amplification was completely linear and there was no need of implementation of the nonlinear effects. However, the linear mechanism is based purely on superposition effects and therefore, it provides the amplification of relative value of the input signal. Laser sources, on the other hand, provide sufficiently high light intensities to modify the optical properties of materials and the light waves interact with each other. This scenario is realized in the present manuscript, where the nonlinear two weakly coupled PCWs are proposed as efficient all-optical digital amplifier. We consider the coupled nonlinear pillar-type PCWs, which consist of the square lattice of the dielectric rods made from a Kerr-type dielectric material located in a free space. One of the main advantages of the pillar type PCW is that it supports one even symmetry mode, which allows for effective single mode operation.

The idea of digital all-optical amplification in this manuscript is based on the phenomenon of band-gap transmission [9,10] in periodic nonlinear media [11,12]. This effect takes place when a frequency of the injected signal is very close to the band edge. In the linear case no transmission occurs, however in nonlinear case above some threshold value of the signal amplitude the propagation of solitons takes place due to nonlinearity of the material. Such a bifurcation could be observed in both spatial and temporal domains [13,14].

V. Jandieri is with the Department of Computer and Electrical Engineering, Free University of Tbilisi, 240 D. Agmashenebeli alley, 0159 Tbilisi, Republic of Georgia (phone: +995-598-112299, e-mail: jandieri@ee.knu.ac.kr). V. Jandieri is a Visiting Scientist at Roma-Tre University, EMLAB Laboratory of Electromagnetic Fields, Department of Engineering, Rome, Italy.

R. Khomeriki is with the Department of Physics, Tbilisi State University, 3 Chavchavadze, 0128 Tbilisi, Republic of Georgia (e-mail: khomeriki@hotmail.com).

Particularly, in waveguide arrays [15,16] one can observe band-gap transmission of self-focusing beams [17], while in optically active media [18] and Bragg gratings [19,20] temporal solitons are created. In principle, the latter two systems could be used for all optical signal amplification, but they do not guarantee a good confinement of the light in the slow light regime, which is necessary condition for enhancement of the nonlinear effects [21]. Therefore, in the present paper we are focusing on the PCWs, where optical losses are negligible. The amplification coefficient is numerically calculated and the applicability and efficiency of the weakly coupled nonlinear PCWs as all-optical digital amplifier is vividly demonstrated. The authors believe that these studies could open the possibility to design novel nonlinear optical devices with the application in all-optical computing systems.

## II. FORMULATION OF THE PROBLEM

The effect of all-optical amplification is analyzed based on the weakly coupled symmetric pillar type PCWs consisting of square lattice of the dielectric rods located in a free space. Geometry of the problem is illustrated in Fig.1. The guiding regions are bounded by the upper and lower photonic crystals composed of $N$ periodic layers and $h$ is a period of the structure along the $z$-axis. The width of the waveguide is $w$. The rods parallel to the $y$-axis are made from a Kerr-type dielectric material and the radius of the rods is $r$. There exist TE and TM modes in the two-dimensional PCWs. In the manuscript we consider TE mode having the electric and magnetic field components ($E_y, H_x, H_y$). The TM mode having the field components ($H_y, E_x, E_y$) can be treated in the same way.

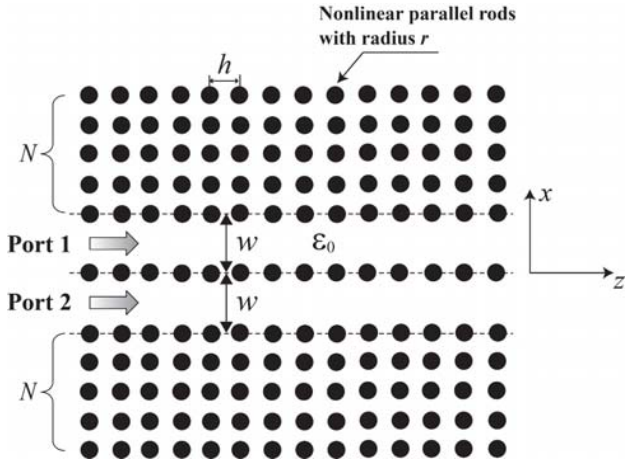

Fig.1. Schematic view of the symmetric coupled nonlinear pillar-type PCWs consisting of square lattice of the dielectric rods periodically distributed along the $z$-axis in a free space, where $h$ is a period of the structure. The rods parallel to the $y$-axis are made from a Kerr-type dielectric material. Number of the layers in both the upper and lower halves of the photonic crystal is denoted by $N$. The signals are launched into the coupled PCWs through Port 1 and Port 2.

Initial are the nonlinear Maxwell's equations, which are written in the following from:

$$\frac{\partial E_y}{\partial z} = -\mu_0 \frac{\partial H_x}{\partial t}, \qquad \frac{\partial E_y}{\partial x} = \mu_0 \frac{\partial H_z}{\partial t}$$
$$\frac{\partial H_z}{\partial x} - \frac{\partial H_x}{\partial z} = \varepsilon_0 n^2 \frac{\partial}{\partial t}\left[ E_y + \chi^{(3)} E_y^3 \right] \tag{1}$$

where $\varepsilon_0$ and $\mu_0$ are the permittivity and permeability of free space, respectively. We introduce the spatially varying index of refraction $n(x,z)$, which we take to be purely real neglecting any extinction due to absorption. $\chi^{(3)}$ is a third-order nonlinear optical susceptibility, which is related to the nonlinear refractive index $n_2$ through the following expression: $\chi^{(3)} = \frac{4}{3} n_2 \varepsilon_0 c$, where $c$ is the speed of light.

In the first-order coupled-mode analysis for the weakly coupled PCWs, the guided field supported by the waveguide system is approximated as follows [22, 23]:

$$E_y(x,z,t) = F(t, z - vt)\Phi(x,z)e^{i(\beta z - \omega_0 t)} + c.c. \tag{2}$$

where $\beta$ is a propagation constant along the $z$-axis, $\omega_0(\beta)$ corresponds to the frequency of the fundamental symmetric $\omega_S(\beta)$ or antisymmetric $\omega_A(\beta)$ modes (indices "$S$" and "$A$" denote "symmetric" and "antisymmetric", respectively), $\Phi(x,z)$ denote the transverse field distribution having the symmetric $\Phi_S(-x,z) = \Phi_S(x,z)$ and the antisymmetric $\Phi_A(-x,z) = -\Phi_A(x,z)$ properties with respect to the $x$-axis and it is periodic function with $z$ with a period $2\pi / h$ [22, 23], $F(t, z - vt)$ is the slowly-varying amplitude and $v_{S,A}$ denotes a group velocity for symmetric and antisymmetric modes, respectively. Evolution of the slowly varying envelope of the pulse electric field amplitude $F_{S,A}(t, z - vt)$ of the symmetric and antisymmetric modes with the change of variables to a moving frame $\xi = z - v_{S,A}t$ satisfies the nonlinear Schrödinger equation:

$$2i\frac{\partial F_{S,A}}{\partial t} + \omega''_{S,A}\frac{\partial^2 F_{S,A}}{\partial \xi^2} + \gamma F_{S,A}\left| F_{S,A} \right|^2 = 0 \tag{3}$$

where $\omega''_{S,A} = \frac{\partial^2 \omega_{S,A}}{\partial \beta^2}$. From Eq.(3) we derive the solitonic solution written in the following form [24]:

$$F_{S,A}(z,t) = \frac{F_{S,A}^0 \exp(i\delta\omega_{S,A}t)}{\cosh\left[ \left( z - v_{S,A}t \right)/\Lambda_{S,A} \right]} \tag{4}$$

with

$$\Lambda_{S,A} = \frac{1}{F_{S,A}^0}\sqrt{\frac{2\omega''_{S,A}}{\gamma}}, \qquad \delta\omega_{S,A} = \frac{1}{4}\gamma\left( F_{S,A}^0 \right)^2 \tag{5}$$

where $\Lambda_{S,A}$ is a soliton width having symmetric or antisymmetric properties, $\delta\omega_{S,A}$ denotes the phase shift of the localized modes due to the nonlinearity of the rods, $F_{S,A}^0$ are soliton amplitudes and the coefficient of nonlinearity is equal to $\gamma = 3\omega_{S,A}\chi^{(3)}\kappa/4$. Note that in the homogeneous nonlinear medium $\kappa = 1$, whereas in case of inhomogeneous nonlinear medium such as pillar type coupled PCWs with the nonlinear rods (Fig.1) $\kappa$ should be defined numerically by comparing Eq.(5) with a shape of the soliton in PCWs.

The principle of all-optical amplification in the coupled nonlinear PCW is as follows: firstly, we properly choose the operating frequency $\omega$ located at the edge of the dispersion curve where no low amplitude propagating modes are excited. Let us assume that the operating frequency $\omega$ is located very close to the frequency at which only the antisymmetric propagating mode is excited in the coupled PCWs. In this case, from Eqs.(4) and (5) it follows that the operating frequency $\omega$ is given as follows:

$$\omega(\beta) = \omega_A(\beta) - \gamma\left(F_A^0\right)^2/4 \qquad (6)$$

We launch a signal at this particular frequency in Port 1 of the coupled PCWs, whereas no signal is injected through Port 2. The input signal could be considered as a linear combination of the symmetric and antisymmetric modes having the same amplitudes. Due to the fact that neither symmetric nor antisymmetric propagating modes are excited at this particular frequency (both of the modes are evanescent modes), the output power of the signal inside the coupled nonlinear PCWs is not observed. Next, additionally a signal with a small amplitude $f$ is injected into the guiding structure through Port 2. Since we are dealing with the weakly coupled PCWs, the signal in the coupled PCWs could be presented as linear combination of the symmetric and antisymmetric modes having different amplitudes: $(1+f)/2$ for symmetric mode and $(1-f)/2$ for antisymmetric mode. The increased power inside the system of the coupled PCWs causes the frequency $\omega$ shift to the higher frequency range due to the nonlinearity of the rods as it follows from Eq.(6). This itself leads to the excitation of the antisymmetric propagating mode, whereas the symmetric mode remains evanescent. As a result, the electric fields of the propagating antisymmetric mode $E_y^A(x,z,t)$ and that of the evanescent symmetric mode $E_y^S(x,z,t)$ are expressed as:

$$E_y^A = \Phi_A(x,z)\frac{F_A^0 \exp[i(\beta_A z - \omega t)]}{\cosh\left[F_A^0(z - v_A t)\sqrt{\gamma/2\omega''}\right]} + c.c. \qquad (7)$$

$$E_y^S = \Phi_S(x,z)F_S^0 \exp(-i\omega t - |\beta_S|z) + c.c. \qquad (8)$$

Finally, the amplification coefficient in the coupled nonlinear PCWs is defined as the ratio of the total output power over the power of the signal injected only in Port 2.

## III. NUMERICAL RESULTS AND DISCUSSIONS

In order to verify the use and applicability of the coupled nonlinear PCWs as all-optical digital amplifier, we numerically study the two coupled symmetric pillar-type PCWs illustrated in Fig.1. Number of the layers of the photonic crystals is $N$=10, which have a common period $h$ along the $z$-axis. The radius of the rod is $r = 0.2h$, its linear refractive index is $n_0 = 3.4$ and $\chi^{(3)}[E_y^0]^2 = 0.56$, where $E_y^0$ is the electric field amplitude of the signal launched into the coupled PCW. The width of the waveguide is $w = 2.0h$ and the length of the photonic crystals is $50h$. In case of pillar-type PCWs, the modes are confined in the guiding region only due to the existence of bandgap region, since the effective index of the guiding region is smaller than that of the cladding region. Under the adjusted parameters, the pillar type photonic crystal has a photonic bandgap for $E$-polarized field ($E_y, H_x, H_y$) in the frequency range $0.277 < h\omega/2\pi c < 0.410$. The dispersion diagrams of the coupled PCWs (Fig.1) are studied based on the coupled-mode formulation [22,23], which we have recently proposed using the first-order perturbation theory taking into account a weak coupling effect. Based on the derived coupled-mode equations propagation constant $\beta h/2\pi$ of the pillar type coupled PCWs (Fig.1) for the symmetric mode and antisymmetric mode versus the normalized frequency $h\omega/2\pi c$ is calculated and it is plotted in Fig.2 by red line and blue lines, respectively. The distributions of the electric field $E_y$ for the symmetric and antisymmetric modes in the cross-section plane $z$=0 are also presented inside the Fig.2. We can see that the fields of the modes in the coupled PCWs are well confined in the guiding region.

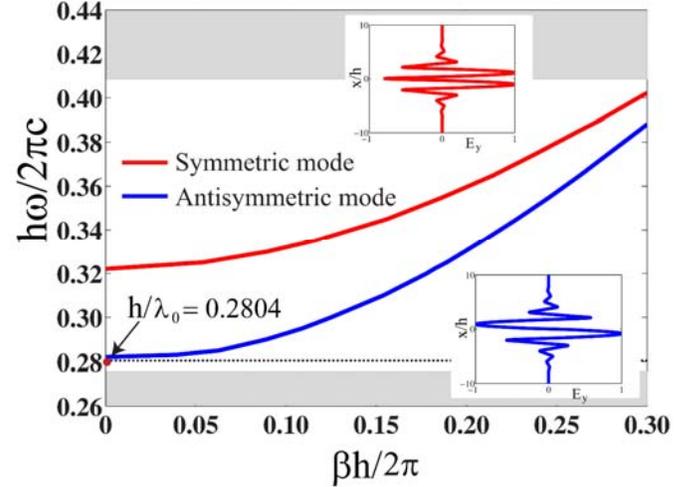

Fig.2. Dispersion curves of the symmetric (red line) and the antisymmetric (blue line) modes of the coupled pillar-type PCWs shown in Fig.1, where $r = 0.2h$, $n_0 = 3.4$ and $w = 2.0h$. The length of the photonic crystals is $50h$. Bandgap region of the upper and lower photonic crystals lies in the frequency range $0.277 < h\omega/2\pi c < 0.410$. The operation frequency is $h\omega/2\pi c = 0.2804$ marked by the red dot. The field distributions of the electric field $E_y$ for the symmetric and antisymmetric modes are presented inside the figure.

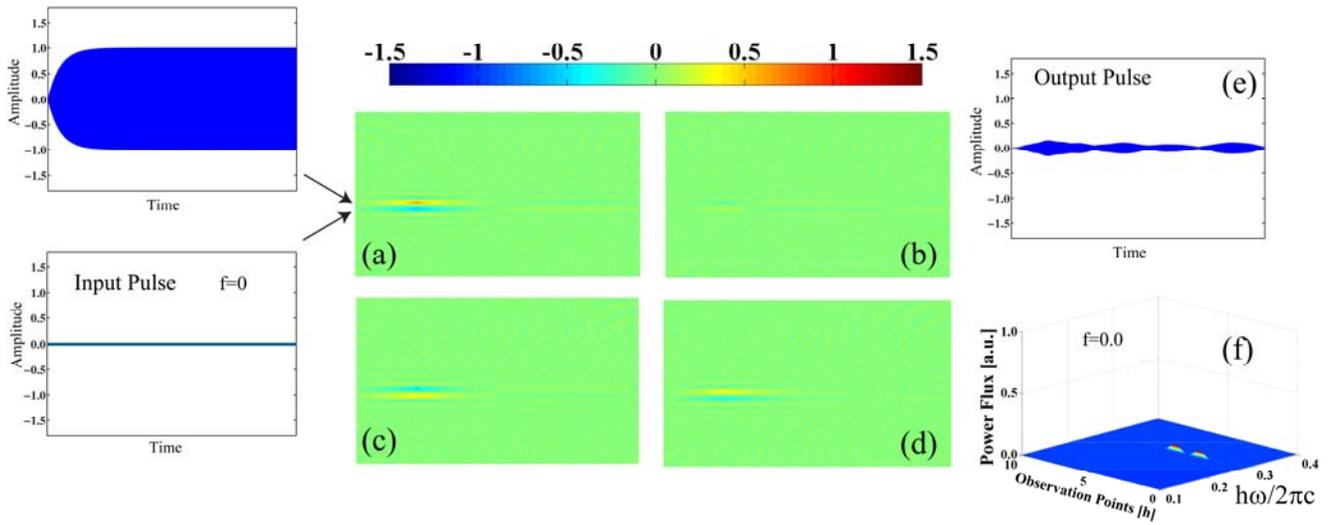

Fig.3. (a)-(d): Near field distributions of the signal propagation inside the coupled nonlinear PCWs (Fig.1), when a continuous input signal is launched through Port 1 and no input signal $f$=0 is injected in Port 2. (e): Amplitude of the electric field $E_y$ of the registered signal by a detector located at a distance $z$=30$h$ inside the coupled PCWs parallel along the $x$-axis. The length of the detector is $\ell$ =10$h$. (f): Power flux of the output signal at a distance $z$=30$h$ inside the coupled nonlinear PCWs.

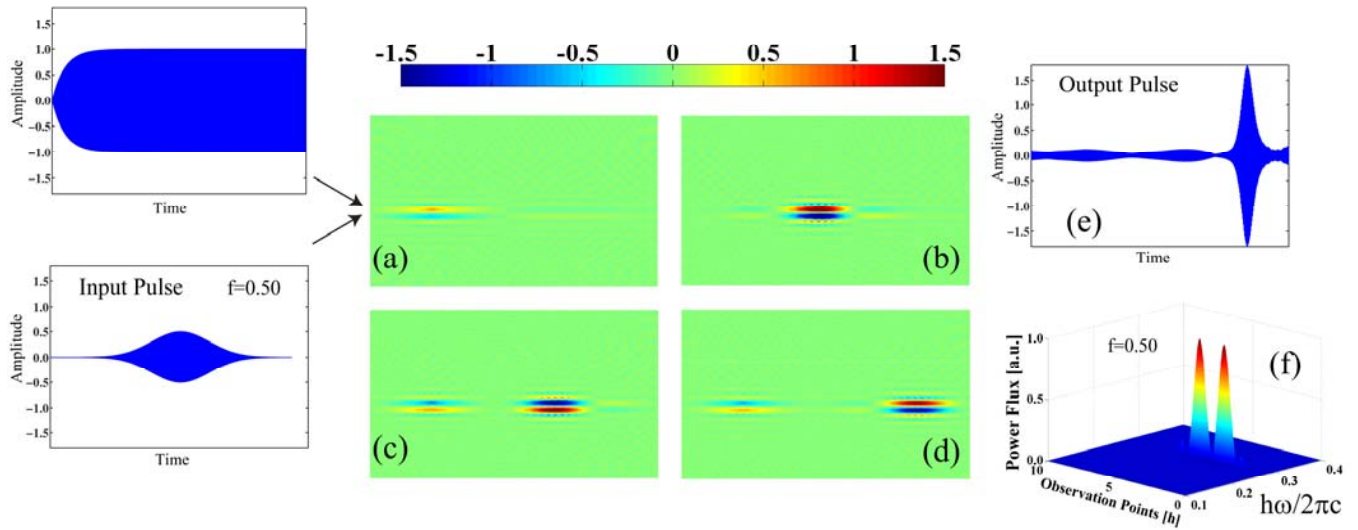

Fig.4. (a)-(d): Near field distributions of the signal propagation inside the coupled nonlinear PCWs (Fig.1), when a continuous input signal is launched through Port 1 and Gaussian pulse with an amplitude $f$=0.5 is injected in Port 2. (e): Amplitude of the electric field $E_y$ of the registered signal by a detector located at a distance $z$=30$h$ inside the coupled PCWs parallel along the $x$-axis. The length of the detector is $\ell$ =10$h$. (f): Power flux of the output signal at a distance $z$=30$h$ inside the coupled nonlinear PCWs.

It should be also noted that no other propagating modes exist in the frequency range $0.277 < h\omega / 2\pi c < 0.410$. Moreover, the structure shown in Fig.1 is characterized without any losses, which makes it a promising candidate for the experimental studies and for the fabrication of all-optical digital amplifier.

All the numerical analyses what follow are conducted at the fixed normalized frequency $h\omega / 2\pi c = 0.2804$ marked by the red dot in Fig.2. It should be emphasized that it is located at the edge of the dispersion curve of the antisymmetric mode (blue line) very near to $h\omega_A(\beta=0) / 2\pi c$ and no propagating modes are excited at this frequency for low input amplitudes. Firstly, a continuous signal with the unit amplitude and the normalized frequency $h\omega / 2\pi c = 0.2804$ is launched into the upper PCW through Port 1, whereas no signal $f$=0 is launched in Port 2. Note that $f$ denotes the amplitude of the signal launched in the lower PCW through Port 2. Numerical analyses are conducted based on the FDTD method using the Berenger's PML [25,26]. Figures 3(a)-3(d) demonstrate the near field distributions of the signal propagating inside the coupled nonlinear PCWs. Amplitude of the electric field $E_y$ and the power flux of the output registered signal by a detector located at a distance $z$=30$h$ inside the coupled PCWs parallel along the $x$-axis are presented in Figs.3(e) and 3(f), respectively. The length of the detector is $\ell$ =10$h$. From Fig.3 it follows that the signal is not propagating inside the coupled nonlinear PCWs since both modes are evanescent modes and the registered output power is negligibly small at $h\omega/2\pi c=0.2804$.

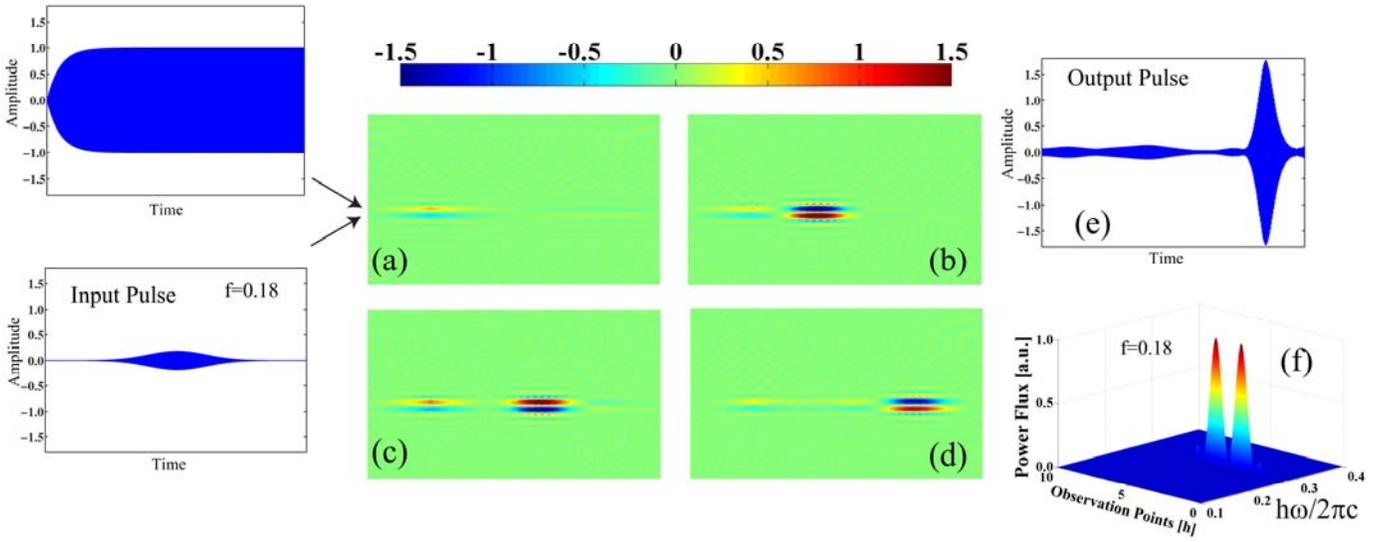

Fig.5. The same as in Fig.4, but at $f$=0.18.

Next, in order to demonstrate the coupled nonlinear PCWs (Fig.1) as an all-optical digital amplifier, we additionally inject a Gaussian pulse in Port 2 with an amplitude $f$=0.5. It should be emphasized that in order to guide an input signal into the coupled PCWs, which supports only one antisymmetric mode marked by blue line in dispersion diagram (Fig.2), the injected signals in Port 1 and Port 2 should be with opposite phases (the signals with the same phases will be completely reflected from the coupled PCWs [8]). An increase of the input power comparison to Fig.3 causes the change of the refractive index of the rods because of Kerr nonlinearity and it leads to the shift of the operating frequency. As a result, the antisymmetric mode becomes the propagating mode (symmetric mode is an evanescent mode) and the soliton is guided inside the PCWs as shown in Figs. 4(a)-4(d). It is important to emphasize that in case of pillar-type PCWs the frequency shift is very small, since the modes are strongly confined in the guiding region (Fig.2) and only the small tails of the field around the rods make an influence on the change of the nonlinear refractive index $n_2$ of the rods. Analyzing numerically the soliton formed in the coupled nonlinear PCWs, we have calculated that the nondimensional parameter $\kappa$, which is related to the coefficient of nonlinearity $\gamma$ in Eq.(5), is equal to $\kappa \approx 2.5\cdot10^{-3}$. Our numerical analyses have shown that the pillar-type nonlinear PCWs (Fig.1) support a slow light mode with a group velocity $c/13$, where the denominator represents the slowdown factor [21]. When a pulse travels in a slow waveguide it is compressed, which in turn increases the peak of the electric field intensity carried by the slow light pulse (Fig.4(e)). The slow light enhancements are accounted for by multiplying $\chi^{(3)}$ by the square of the slowdown factor (in our case slowdown factor is equal to 13). The output power flux at $f$=0.5 is also presented in Fig.4(f). Amplification coefficient $r$, which is calculated as a ratio of the total output power over the input power launched through Port 2 is $r=3.1$.

Finally, similar numerical analyses as in Fig.4 are conducted when the amplitude $f$ of the injected signal in Port 2 is $f$=0.18. The results are demonstrated in Fig.5. Numerical studies have shown that the amplification coefficient $r$ increases inversely proportional to the amplitude $f$ and it is equal to $r=9.5$ at $f$=0.18. Our analysis vividly demonstrate the applicability and effectiveness of the coupled pillar type nonlinear PCWs (Fig.1) as an all-optical digital amplifier. Another interesting point of worth mentioning is that the total output power flux of the signal registered inside the coupled nonlinear PCW does not depend on the input power launched in Port 2 and it is the same at various values of the amplitude $f$. This effect could be vividly seen when comparing the output power fluxes in Fig.4f (at $f$=0.5) and Fig.5f (at $f$=0.18), respectively. This feature uniquely characterizes the coupled nonlinear PCWs and it is never observed in linear case when the nonlinear effects are not taken into account.

## IV. CONCLUSION

In the manuscript, for the first time, we have proposed a weakly coupled nonlinear pillar-type PCWs as a realistic model for all-optical digital amplifier. The deep physical insight has been given to the realization of all-optical amplification effect of the light pulse in the coupled nonlinear PCWs. The numerical examples have shown the validity of the theoretical analysis. The amplification coefficient has been calculated for the various amplitudes of the input signal launched into the coupled nonlinear PCWs. The results have demonstrated the effectiveness of the proposed structures as all-optical digital amplifier. We believe that one of the promising practical applications of the proposed structure could be optical transistors in all-optical computing systems. Being able to precisely control one light beam using another is crucially important for the development of the optical transistors capable of performing complex light-controlling in all-optical circuits.

ACKNOWLEDGEMENT

R. Khomeriki acknowledges financial support from Georgian SRNSF (grant No 30/12).